\documentclass[aps,prb,final,twocolumn,superscriptaddress,showkeys,showpacs]{revtex4}

\usepackage{graphicx}% Include figure files
\usepackage{dcolumn}% Align table columns on decimal point
\usepackage{bm}% bold math
\usepackage{amsmath}
\usepackage{amssymb}

\newcommand\abbwidth{0.9\linewidth}

\begin{document}
\title{Terahertz photoresponse of a quantum Hall edge-channel diode}

\author{Christian Notthoff}
\email[]{christian.notthoff@uni-due.de}
\affiliation{Fachbereich Physik and CeNIDE, Universit\"at Duisburg-Essen, Lotharstr.1, Duisburg 47048, Germany}
\author{Kevin Rachor}
\author{Detlef Heitmann}
\affiliation{Institut f\"ur Angewandte Physik, Universit\"at Hamburg, Hamburg 20355, Jungiusstr. 11, Germany}
\author{Axel Lorke}
\affiliation{Fachbereich Physik and CeNIDE, Universit\"at Duisburg-Essen, Lotharstr.1, Duisburg 47048, Germany}

\date{\today}

\begin{abstract}
The Teraherz (THz) photoresponse of a two-dimensional electron gas in the quantum Hall regime is investigated. We use a sample structure which is topologically equivalent to a Corbino geometry combined with a cross-gate technique. This quasi-Corbino geometry allows us to directly investigate the THz-induced transport between adjacent edge-states, thus avoiding bulk effects. We find a pronounced photo voltage at zero applied bias, which rapidly decreases when an external current bias is applied. The photo voltage and its dependence on the bias current can be described using the model of an illuminated photodiode, resulting from the reconstruction of the Landau bands at the sample edge. Using the sample as a detector in a Fourier transform spectrometer setup, we find a resonant response from which we extract a reduced effective cyclotron mass. The findings support a non-bolometric mechanism of the induced photo voltage and the proposed edge-channel diode model.
\end{abstract}

% insert suggested PACS numbers in braces on next line
% insert suggested keywords - APS authors don't need to do this

\preprint{revision 1}
% \maketitle must follow title, authors, abstract, \pacs, and \keywords
\maketitle
% \texttt{http://ees.elsevier.com/physe}.
\section{Introduction}
It has been found almost 30 years ago that quantum Hall systems can be excited effectively by Terahertz (THz) radiation, because the energy gap between the Landau levels is comparable to the photon energy (3~THz$\hat{=}$12.4~meV). First photoconductivity investigations using GaAs/AlGaAs heterostructures, patterned in a Hall bar, showed a strong photoresponse $\Delta R_{xx}$ of the longitudinal magnetoresistance $R_{xx}$ under THz radiation\cite{SSC50,APL40,surf142}. In these studies, two contributions to the photoresponse were found: a non-resonant signal as well as a cyclotron resonant (CR) component. Furthermore, it was observed that the magnetic field dependence of $\Delta R_{xx}$ is similar to that of the temperature derivative of the longitudinal resistance $R_{xx}$ (see e.g. Refs.\onlinecite{SSC50,JAP}). Today it is well established that the non-resonant part is caused by a lattice heating of the whole sample, whereas the CR signal arises form electron gas heating due to the cyclotron absorption inside the two-dimensional electron gas (2DEG). More recently, Hirakawa et al.\cite{PRB63} have observed a deviation of the photoresponse from the electron gas heating model at filling factors $\nu>2$ and low bias currents. They suggested that the deviation from the heating model arises from the non-equilibrium edge-channel transport, which  affects the photoresponse. However, the effect of edge-channel transport on the photoresponse is not yet fully understood.

Recently, W\"urtz et al.\cite{alida} reported on the development and application of a sample geometry which allows a direct measurement of the transport between adjacent edge-channels at the sample boundaries, where the two edge-channels are separated only by one incompressible strip. This allows for the investigation of the transport properties across the incompressible strip using four-point resistance measurements\cite{alida,PRLalida,deviat1,deviat2}.\\
In this paper we will present THz photoresponse measurements on a quasi-Corbino sample. The experimental data from our DC and THz measurements suggest that the origin of the photoresponse is a photovoltaic effect, rather than a bolometric change in conductivity. We show that the photo voltage arises from a photocurrent, generated inside the incompressible edge strip, which acts as an illuminated "edge-channel diode". Furthermore, spectrally resolved measurements reveal a reduced effective cyclotron mass, which indicates that the photo voltage originates from the edge of the sample and supports the photodiode model to describe the photo voltage.

%Here we present THz photoresponse measurements on such a device and show that the origin of the observed photovoltaic effect, is a photocurrent, generated inside the incompressible edge strip, which acts as an illuminated "edge-channel diode". \\
%This paper is organized as follows: Section \ref{sec:expsetup} gives a description of the sample geometry and the experimental setup for the photoresponse measurements. A brief resume of the results presented in Ref.\cite{alida} is given in Sec. \ref{sec:model}. Finally the experimental results are presented  in Sec. \ref{sec:expres} and discussed in terms of a photodiode model.
\section{Experimental Setup}\label{sec:expsetup}
A two dimensional electrongas (2DEG) is formed in a MBE (molecular beam epitaxy) grown GaAs/AlGaAs heterostructure, consisting of a 1~$\mu$m~GaAs buffer, 20~nm undoped AlGaAs, followed by a Si $\delta$-doping layer, and again 5~nm~AlGaAs, covered by a 40~nm superlattice, and 5~nm~GaAs cap layer. The carrier concentration and the mobility are about $n= 4.0\times10^{11}$cm$^{-2}$ and $\mu=800.000~$cm$^2$/Vs, respectively, at 4.2~K. The sample structure is defined by standard photolithography and wet chemical etching. The gate electrode consists of 50~nm thermally evaporated Au. Ohmic contacts are provided by alloyed AuGe/Ni/AuGe~$(88:12)$ pads.
%The sample is fabricated from a MBE grown GaAs/AlGaAs heterostructure, consisting of a 1~$\mu$m~GaAs buffer, 20~nm undoped AlGaAs, followed by a $\delta$-doping layer, and again 5~nm~AlGaAs, covered by a 40~nm superlattice, and 5~nm~GaAs cap layer. The carrier concentration and the mobility are about $n= 4.0\times10^{11}$cm$^{-2}$ and $\mu=800000~$cm$^2$/Vs, respectively at 4.2~K. The sample structure is defined by standard photolithography and wet chemical etching. The gate electrode consists of 50~nm thermally evaporated Au. Ohmic contacts are provided by alloyed AuGe/Ni/AuGe~$(88:12)$ pads.
\begin{figure}[htb]
  \begin{center}\leavevmode
    \includegraphics[width=\abbwidth]{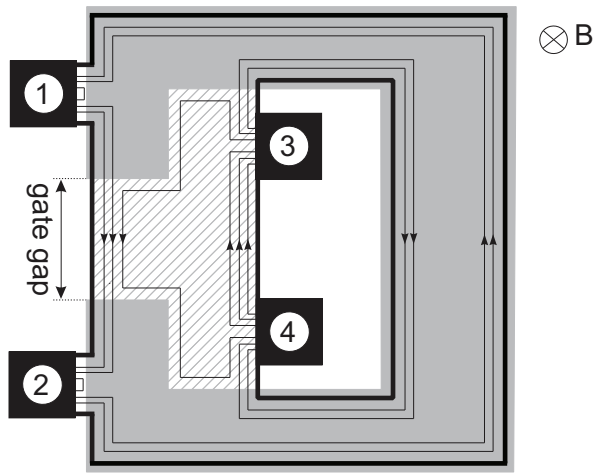}
    \caption{Sketch of the sample geometry. Contacts are positioned along the inner and outer edges of the ring-shaped mesa. The shaded area represents the Au gate. The hatched area is the ungated region of the 2DEG. Arrows indicate the direction of electron drift motion in the edge-channels for the outlined configuration: filling factor $\nu=3$ in the ungated region and $g=2$ under the gate (see also Ref. \onlinecite{alida}).} \label{fig:sample_setup}
  \end{center}
\end{figure}\\
%The sample geometry (see Fig.~\ref{fig:sample_setup}) is equivalent to the geometry used in Refs.\cite{alida,PRLalida,deviat1,deviat2}. 
Figure \ref{fig:sample_setup} schematically shows the topology of the sample, consisting of a ring-shaped (Corbino) mesa and a horse-shoe shaped gate. The ring-shaped mesa is shown by the thick outline and the gate electrode is indicated by the grey shaded area. In the quantum Hall regime and at zero gate voltage, edge-channels develop along the sample boundaries, which are well separated by the semi-insulating bulk in the inner region. Therefore, the present quasi-Corbino geometry has two independent sets of edge-channels, called inner and outer edge-states in the following. Applying a suitable negative gate voltage, the inner edge-state can be redirected to run along the outer edge-state in the gap region (see thin lines with arrows in Fig.~\ref{fig:sample_setup}). Whereas the outer edge-states are connected to the outer Ohmic contacts 1 and 2, the innermost edge-states are connected to the interior contacts 3 and 4. This situation is shown in Fig.~\ref{fig:sample_setup}, where the filling factors have been adjusted to $\nu=3$ in the ungated region and $g=2$ under the gate. When the bulk region of the 2DEG under the gate is in its insulating state (integer filling factor and sufficiently low temperatures) current transport between inner and outer Ohmic contacts is possible only by charge equilibration among neighboring channels in the gap region. The present geometry therefore allows for a direct investigation of transport across the incompressible strip that separates the compressible edge-channels within the gate gap\cite{alida,PRLalida,deviat1,deviat2}. Furthermore, as we will show in the following, it makes it possible to investigate the THz-induced transport between two edge-states, without any bulk effects.\\

In the present experiment, we record the photo voltage $\Delta V$ as a function of the bias current. A constant current $I$ is applied between contacts 1 and 3 (see Fig.~\ref{fig:sample_setup}) and the non-local voltage at contacts 2 and 4 is measured using a high-impedance ($\geq 100~\text{M}\Omega$) voltage preamplifier. The Hg lamp of a Bruker IFS113v Fourier-transform spectrometer (FTS), a black poly filter, and a $23\mu$m Mylar beam splitter is used as a THz light source. We perform two types of experiments. In the integrated mode we stop the mirror scanner of the FTS, chop the THz radiation at 20~Hz with a chopper inside the beam-path and measure the photo voltage in lock-in technique. In a second mode we use, without chopper, the spectrally resolved mode of the FTS. A system of mirrors, and polished stainless steel tubes, acting as oversized waveguides, guides the THz radiation to the sample, mounted inside a He$^3$ cryostat with a base temperature of 300~mK. The He$^3$ system it self is located in a He$^4$ cryostat with a superconducting magnet, to apply fields up to $B=8$~T, where B is perpendicular to the surface normal of the 2DEG.
%and a chopper at 20~Hz inside the beam-path is used as a modulated THz light source. A system of mirrors, and polished stainless steel tubes, acting as oversized waveguides, guides the THz radiation to the sample, mounted inside a He$^3$ cryostat with a base temperature of 300~mK. The He$^3$ system it self is located in a He$^4$ cryostat with a superconducting magnet, to apply fields up to $B=8$~T. 
%%%%%%%%%%%%%%%%%%%%%%%%%%%%%%%%%%%%%%%%%%%%%%%%
%%%%%%%%%%%%%%%%%%%%%%%%%%%%%%%%%%%%%%%%%%%%%%%%
\section{Diode model}\label{sec:model}
%%%%%%%%%%%%%%%%%%%%%%%%%%%%%%%%%%%%%%%%%%%%%%%%
%%%%%%%%%%%%%%%%%%%%%%%%%%%%%%%%%%%%%%%%%%%%%%%%
In this section we will give a brief description of the findings in Ref.\onlinecite{alida} and the interpretation of the experimental observations in terms of a backward diode model, based on the picture of edge-state reconstruction developed by Chklovskii et al.\cite{PRB46}.
\begin{figure}[htb]
  \begin{center}\leavevmode
    \includegraphics[width=\abbwidth]{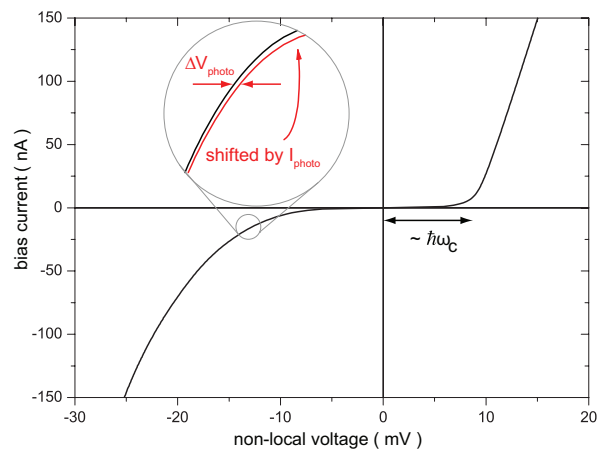}%HH_6_11_Ug-140mV_B=6_066T_dUcalc_aus_Iphoto_Urr_neu
    \caption{(Color online) $I-V$ trace across the Landau gap, measured at 300~mK and $B=5.935$~T. The filling factors are chosen to $\nu=3$ and $g=2$. The inset illustrates the occurrence of a photo voltage $\Delta V_{photo}$ in a four-probe experiment due to a photo current $I_{photo}$ induced inside the incompressible edge strip in the framework of a photodiode model. The red curve shows the same data as the black curve, but shifted by a photo-current of $3$ pA. Note that the data was taken by four-probe $V(I)$ measurements and that the axes have been exchanged for better comparison with standard diode characteristics.} \label{fig:DC_Urr_Isd}
  \end{center}
\end{figure}\\
Figure \ref{fig:DC_Urr_Isd} shows a typical nonlinear current-voltage ($I-V$) trace measured in a four-probe geometry for the filling factors $\nu=3$ in the ungated region and $g=2$ under the gate at a temperature of 300~mK (see also Fig.~\ref{fig:sample_setup}). The measured $I-V$ trace resembles those of a backward diode.
\begin{figure}[htb]
  \begin{center}\leavevmode
    \includegraphics[width=\abbwidth]{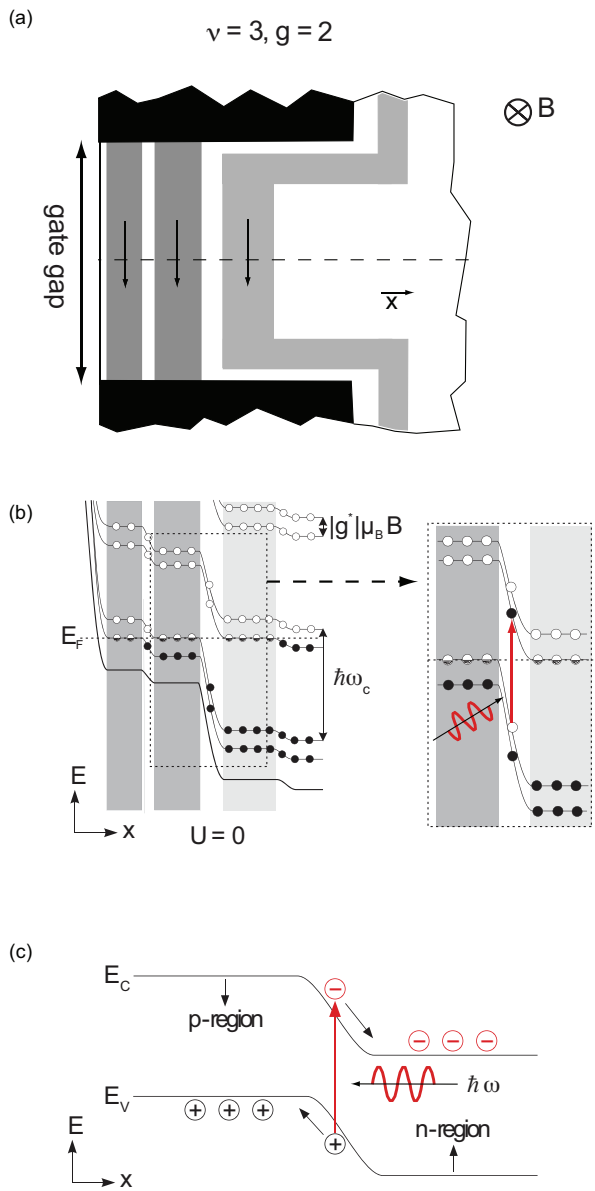}%edge_potential.
    \caption{(Color online) Application of the edge-reconstruction model to the experimental setup. (a) Gate gap region with compressible (shaded) and incompressible (light) strips. (b, left) Edge potential as derived from the compressible and incompressible strips model in the case of a constant Fermi energy in the gate gap region ($V=0$). Filled circles indicate occupied and open ones unoccupied states. (b,right) Illustrates a photo excitation process inside the most inner incompressible strip. (c) Text book energy level scheme of a photodiode.} \label{fig:potential}
  \end{center}
\end{figure}\\
Indeed, in the picture of edge-state reconstruction (taking into account screening effects among electrons\cite{PRB46}), transport across an incompressible edge strip can be described in a manner comparable to the transport across the depletion layer in a p-n backward diode\cite{alida,PRLalida,deviat1,deviat2,PhysicaB211}. Figure \ref{fig:potential} (a) illustrates the spatial distribution of the edge-channels in the gap region for the filling factors $\nu=3$ in the ungated region and $g=2$ under the gate (see also Fig.~\ref{fig:sample_setup}). In Fig.~\ref{fig:potential} (b) we have sketched the edge-state reconstruction at zero bias current\cite{alida,PRB46} along the section indicated by the dashed line in Fig.~\ref{fig:potential} (a). As depicted in the right panel of Fig.~\ref{fig:potential} (b), the edge-state reconstruction is similar to the energy level scheme of a p-n diode sketched in Fig.~\ref{fig:potential} (c). Furthermore, a characteristic onset voltage of the order of the Landau gap ($\hbar \omega_c/e$) is observed at positive bias currents and an almost constant differential resistance at biases exceeding this onset voltage. For reverse bias the $I-V$ trace shows a nonlinear behavior, but no distinct onset voltage can be observed. Under forward bias no current will flow until almost flatband conditions are reached, because the thermal energy ($\approx0.03$~meV) is much smaller than the built-in potential barrier ($\approx\hbar \omega_c=9$~meV). This explains the abrupt current onset at $eV\approx\hbar\omega_c$ in Fig.~\ref{fig:DC_Urr_Isd}. For bias above  $eV\approx\hbar\omega_c$ the differential resistance almost vanishes. For reverse bias, tunneling through the incompressible edge strip sets in, which also leads to a non-linear $I-V$ characteristic, however without a distinct onset voltage. A more detailed discussion of the transport characteristics of the present sample topology under dark condition can be found in Refs. \onlinecite{alida} and \onlinecite{deviat2}.
%%%%%%%%%%%%%%%%%%%%%%%%%%%%%%%%%%%%%%%%%%%%%%%%
%%%%%%%%%%%%%%%%%%%%%%%%%%%%%%%%%%%%%%%%%%%%%%%%
\section{Experimental results}\label{sec:expres}
%%%%%%%%%%%%%%%%%%%%%%%%%%%%%%%%%%%%%%%%%%%%%%%%
%%%%%%%%%%%%%%%%%%%%%%%%%%%%%%%%%%%%%%%%%%%%%%%%
\subsection{Spectrally integrated measurements}
\begin{figure}[htb]
  \begin{center}\leavevmode
  \includegraphics[width=\abbwidth]{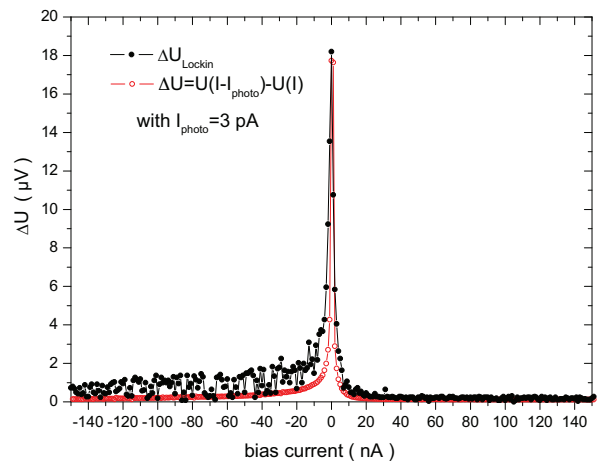}%HH_6_11_Ug-140mV_B=6_066T_Lockin_vgl_dUcalc_aus_Iphoto
    \caption{(Color online) Four-terminal photo voltage $\Delta V$ (filled black circles) measurement in lock-in technique at 300~mK and $B=5.935$~T for the adjusted filling factor combination $\nu=3$ and $g=2$. The red open circles indicate the calculated photo voltage according to equation (\ref{eq:def_U_potodiode}), where a photo current of $I_{photo}=3~$pA and the $I-V$ trace measured in the dark shown in Fig.~\ref{fig:DC_Urr_Isd} is used.} \label{fig:PC_Isd}
  \end{center}
\end{figure}

Figure \ref{fig:PC_Isd} shows the measured photo voltage $\Delta V$ (filled circles) of the device.
We observe a pronounced maximum of $\Delta V=18~\mu$V at zero bias current and a strongly decreasing signal for higher bias currents, in either polarity. Furthermore, the experimentally obtained photo voltage exhibits an asymmetry between positive and negative bias currents. With increasing positive current, the photo voltage decreases very fast and no signal can be observed above $I\approx 10-20$~nA, whereas at negative bias a small signal remains even at a bias current of $I=-140$~nA. The occurrence of a photovoltage at zero bias current is striking and clearly shows that the signal is not caused by a bolometric effect (or any other effect that merely changes the resistance of the sample). Rather, there must be a mechanism which converts the THz illumination into a separation of charge. Absorption in the bulk is not a likely explanation, since it would not lead to the breaking of symmetry required for the observation of a distinct voltage. Furthermore, an effect of illumination of the ohmic contacts can be ruled out, because only the gate gap region is irradiated.\\
To quantitatively explain the photo voltage of a quasi-Corbino sample, we take the model discussed above one step further and assume that the reconstructed edge (see Fig. \ref{fig:potential} (b,c)) is comparable to a diode not only under dark conditions but also under illumination with THz radiation. The photo excitation process inside an incompressible edge strip at zero bias current is sketched on the right panel of Fig.~\ref{fig:potential} (b). Because of the built-in electric field in the incompressible edge strip, photo-excited electron-hole pairs become spatially separated, which causes a negative photocurrent even at zero bias current in complete analogy to the case of a photodiode\cite{Sze} (see Fig.~\ref{fig:potential} (c)). As depicted in the inset of Fig.~\ref{fig:DC_Urr_Isd} we therefore expect that the $I-V$ trace is shifted along the current axis to negative values under illumination. For $I=0$, this shifted $I-V$ characteristic reveals it self as the open circuit photo voltage $\Delta V_{photo}(I=0)$. Assuming a constant photocurrent $I_{photo}$\cite{Sze}, the photo voltage is then given by
\begin{equation}\label{eq:def_U_potodiode}
\Delta V_{photo}(I):=V(I-I_{photo})-V(I)~,
\end{equation}
at any bias current $I$ (see inset in Fig.~\ref{fig:DC_Urr_Isd}). To calculate the photo voltage from the $I-V$ trace according to equation (\ref{eq:def_U_potodiode}) we have to know the photocurrent. But knowing the zero bias resistance $R_{diode}$ of the diode, it is straight forward to derive the photocurrent $I_{photo}$ from the photo voltage $\Delta V(I=0)$:
\begin{equation}\label{eq:calc_I_photo}
I_{photo}=\frac{\Delta V_{photo} }{ R_{diode}}=\frac{18~\mu\text{V}}{6~\text{M}\Omega} =3~\text{pA}~,
\end{equation}
where $R_{diode}\approx 6~\text{M}\Omega$ is obtained from the slope of the $I-V$ trace around $I=0$ (see Fig.~\ref{fig:DC_Urr_Isd}).\\
Figure \ref{fig:PC_Isd} shows the photo voltage, calculated from the $I-V$ trace in the dark shown in Fig.~\ref{fig:DC_Urr_Isd}, according to equations (\ref{eq:def_U_potodiode}) and (\ref{eq:calc_I_photo}) (open circles), compared to the experimentally observed photo voltage (closed circles). The good agreement between the calculated and the measured photo voltage, without any fit parameter, strengthens the picture discussed above that the transport across a single incompressible edge strip can be described within a simple backward diode model, under dark condition as well as under illumination with THz radiation. It should be pointed out that already 20 years ago a similar mechanisem was proposed by Liu et al. \cite{APL55} for Cobion disks with a gate-induced potential step in the bulk of the sample.\\
Furthermore, the observed asymmetry in Fig.~\ref{fig:PC_Isd} is also reproduced by the calculated photo voltage. It results from the fact that for bias currents $|I|>10$~nA the differential resistance is higher for negative polarity than for positive polarity. Accordingly, a shift in current (induced by the photo-generated carriers) translates into a smaller voltage for $I>10$ nA than for $I\lesssim-20$ nA. We would like to note that we measure very similar behavior at filling factors $\nu=4$ and $g=2$ (not shown).
%The picture of a edge-channel diode is further supported by the results of Liu et al. \cite{APL55}. They utilized a carrier density discontinuity, induced by a front gate, and find a diode like $I-V$ characteristic if the gate is biased such that the Fermi level at the gate edge must cross adjacent Landau levels. Under illumination they observe a photovoltaic effect comparable to the present results above.
\subsection{spectrally resolved measurements}
For spectrally resolved measurements we remove the chopper, turn on the moving mirror of the FTS and use the sample as a photo detector to record the intensity of the modulated light. In the spectral range of interest, the light source and all optical components have a very weak dependence on wavelength, so that the recorded signal will be dominated by the spectral response of the sample it self (for a similar experiment to record the bolometric response in the quantum Hall effect, see Ref. \onlinecite{EP2DS}).
\begin{figure}[htb]
  \begin{center}\leavevmode
    \includegraphics[width=\abbwidth]{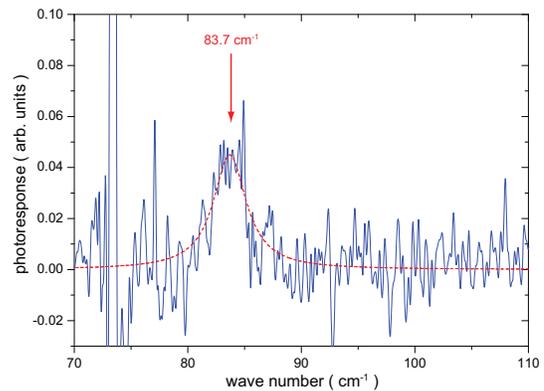}%HH_6_11_Ug-141mV_B=6_14T_Spektral
    \caption{(Color online) Spectrally resolved photo voltage (solid line) at 300~mK and $B=6.008$~T for the adjusted filling factor combination $\nu=3$ and $g=2$. The photo voltage is measured between contacts 2 and 6 at zero bias current. The dashed line shows the best Lorentz fit to the data.} \label{fig:spectral}
  \end{center}
\end{figure}
Figure \ref{fig:spectral} shows the spectral response of the sample, obtained by feeding the open circuit voltage between contacts 1 and 3 into the FTS and processing 800 co-added interferograms. A weak single resonance is observed, where the signal amplitude is about a factor of 2000 weaker than the photoconductivity signal on a sample from the same wafer but in standard Corbino topology (see Ref.\onlinecite{EP2DS}). As indicated by the dashed line, the resonance can be fitted by a Lorentzian profile with a resonance position at $\overline{\nu}=83.7~\text{cm}^{-1}$. The linewidth of $3.53~\text{cm}^{-1}$ is comparable to the width expected from the bolometric, photoconductivity signal ($\approx 2~\text{cm}^{-1}$) in standard Corbino topology and the width deduced from the transport mobility and the carrier density of the sample (see e.g. Ref.\onlinecite{SURF58}). From the resonance position and the applied magnetic field we can deduce an effective cyclotron mass of $m_{cr}=0.0670~m_e$, where $m_e$ is the free electron mass. This mass, surprisingly, is somewhat smaller than the cyclotron mass commonly observed in GaAs/AlGaAs heterostructures at high magnetic fields ($m_{cr}\approx 0.069~m_e$)\cite{STC66,PR122,PRB53,bnp}. It is also smaller than the cyclotron mass of $0.0692~m_e$ which has been observed for the 2D bulk, bolometric, photoconductivity signal on a sample from the same wafer but in standard Corbino topology (see Ref.\onlinecite{EP2DS}).\\
The observation of a reduced effective mass further supports the conclusion that the observed photo voltage does not arises from the 2D bulk cyclotron resonance as the bolometric contribution. A simple explanation for the reduction of the effective mass compared to the 2D cyclotron mass is that the electrons experience an additional confinement potential due to the depletion field inside the incompressible edge strips. Such an additional confinement will shift the observed resonance to higher frequencies, resulting in a reduced effective cyclotron mass (see also Refs.\onlinecite{PRB63,Lorke}). Therefore, the observation of a reduced effective mass supports the picture that the photo voltage is caused by a photocurrent generated inside the incompressible edge strip between two adjacent edge-channels.\\
In contrast to the spectrally resolved photoresponse of samples in standard Corbino topology (see Ref.\onlinecite{dissIch}) and samples in hall-bar topology (see Ref.\onlinecite{PRB63}), where a asymmetric broadening at filling factors $\nu>2$ is observed, here we observe only one symmetric resonance. In fact, we identify the resonance in Fig. \ref{fig:spectral} as the high-energy contribution, which leads to the asymmetry observed in Refs.\onlinecite{PRB63,dissIch}. The low-energy contribution, which is caused by bolometric effects in the bulk is not observed in the present experiment, which is mostly sensitive to the edge excitation.
%%%%%%%%%%%%%%%%%%%%%%%%%%%%%%%%%%%%%%%%%%%%%%%%
%%%%%%%%%%%%%%%%%%%%%%%%%%%%%%%%%%%%%%%%%%%%%%%%
\section{Conclusion}
%%%%%%%%%%%%%%%%%%%%%%%%%%%%%%%%%%%%%%%%%%%%%%%%
%%%%%%%%%%%%%%%%%%%%%%%%%%%%%%%%%%%%%%%%%%%%%%%%
In conclusion, we have used a quasi-Corbino sample geometry to study the photoresponse of a two-dimensional electron gas in the quantum Hall regime. The sample geometry allows for the investigation of THz-induced transport across only one incompressible strip. A photo voltage is observed with a characteristic dependence on the bias current which shows that the origin of the photo voltage is clearly a THz-induced photo current and not a change in conductivity, which is the dominating mechanism in Hall bar samples. The spectrally resolved measurements give a reduced effective mass, which further supports that the photo current originates from the edge of the sample. Together with the good agreement between the experimental observed photo voltage and the one calculated from a photodiode model, we can further conclude that the photo current is generated inside the incompressible edge strips, which acts as an illuminated "edge-channel diode".\\
Furthermore, the model of an illuminated "edge-channel diode" offers an explanation for the asymmetric broadening observed in spectrally resolved photoresponse measurements on samples in standard Corbino topology (see Ref.\onlinecite{dissIch}) as well as in hall-bar topology (see Ref.\onlinecite{PRB63}).
\begin{acknowledgments}
We wish to thank A. W\"urtz for the preparation of the sample and acknowledge financial support by the Deutsche Forschungsgemeinschaft (DFG,project no. LO 705/1-3 and SFB 508).
\end{acknowledgments}

\end{document}